\documentclass[aps,pre,superscriptaddress,citeautoscript,twocolumn,reprint,longbibliography,floatfix]{revtex4-2}
\usepackage{graphicx,amssymb,amsmath,epsf,bm,cprotect,comment,physics}
\usepackage{color,soul}
\epsfclipon

\newcommand{\unit}[1]{\,\mathrm{#1}}

\newcommand{\vO}{\vu{\Omega}}
\newcommand{\vt}{\vu{t}}

\renewcommand{\eqref}[1]{Eq.\,(\ref{#1})}
\newcommand{\eqsref}[1]{Eqs.\,(\ref{#1})}
\newcommand{\pref}[1]{(\ref{#1})}
\newcommand{\figref}[1]{Fig.\,\ref{#1}}
\newcommand{\figpref}[2]{Fig.\,\ref{#1}(#2)}

\begin{document}

\title{Approach and rotation of reconnecting topological defect lines in liquid crystal}

\author{Yohei Zushi}
\affiliation{Department of Physics,\! The University of Tokyo,\! 7-3-1 Hongo,\! Bunkyo-ku,\! Tokyo 113-0033,\! Japan}%

\author{Cody D. Schimming}
\affiliation{Theoretical Division and Center for Nonlinear Studies, Los Alamos National Laboratory, Los Alamos, New Mexico 87545, USA}

\author{Kazumasa A. Takeuchi}
\email{kat@kaztake.org}
\affiliation{Department of Physics,\! The University of Tokyo,\! 7-3-1 Hongo,\! Bunkyo-ku,\! Tokyo 113-0033,\! Japan}%
\affiliation{Institute for Physics of Intelligence,\! The University of Tokyo,\! 7-3-1 Hongo,\! Bunkyo-ku,\! Tokyo 113-0033,\! Japan}

\date{\today}

\begin{abstract}
Topological defects are a universal concept across many disciplines, such as crystallography, liquid-crystalline physics, low-temperature physics, cosmology, and even biology. In nematic liquid crystals, topological defects called disclinations have been widely studied. 
For their three-dimensional (3D) dynamics, however, only recently have theoretical approaches dealing with fully 3D configurations been reported. 
Further, recent experiments have observed 3D disclination line reconnections, a phenomenon characteristic of defect line dynamics, but detailed discussions were limited to the case of approximately parallel defects. 
In this study, we focus on the case of two disclination lines that approach at finite angles and lie in separate planes, a more fundamentally 3D reconnection configuration. 
Observing and analyzing such reconnection events, we find the square-root law of the distance between disclinations and the decrease of the inter-disclination angle over time. 
We compare the experimental results with theory and find qualitative agreement on the scaling of distance and angle with time, but quantitative disagreement on distance and angle relative mobilities. 
To probe this disagreement, we derive the equations of motion for systems with reduced twist constant and also carry out simulations for this case. These, together with the experimental results, suggest that deformations of disclinations may be responsible for the disagreement.
\end{abstract}

\maketitle

\section{Introduction} 

Where there is order, its mismatch exists. 
Topological defects are prime examples of mismatching order and are omnipresent across many materials \cite{Chaikin.Lubensky-Book2000, Nakahara-Book2003}, including crystals \cite{Chaikin.Lubensky-Book2000},  liquid crystals \cite{Chaikin.Lubensky-Book2000,Chandrasekhar-Book1992, deGennes.Prost-Book1995}, superfluids \cite{Zurek-PhysicsReports1996}, spacetime in cosmology \cite{Vilenkin.Shellard-Book2000}, and living things \cite{Doostmohammadi.etal-NatCommun2018, Doostmohammadi.Ladoux-TrendsinCellBiology2022}. Despite various orders depending on systems, corresponding topological defects are expected to share common behaviors. For example, line-shaped defects often experience reconnections \cite{Vilenkin.Shellard-Book2000, Bewley.etal-PNAS2008, Paoletti.etal-PRL2008, Paoletti.etal-PhysicaD2010, Fonda.etal-PNAS2014, Minowa.etal-SA2022, Serafini.etal-PRL2015, Serafini.etal-PRX2017, Chuang.etal-S1991, Zushi.Takeuchi-PNAS2022}, in which two defect lines approach, collide, exchange endpoints, and separate. 

Liquid crystals, especially nematic liquid crystals, are a material where topological defects have been heavily studied because of their optical properties and controllability. 
Nematic liquid crystals are typically composed of building blocks with elongated shape which tend to align with each other. 
They are characterized by a unidirectional order represented by a unit vector $\vu{n}$ with apolar nature ($\vu{n}\equiv -\vu{n}$) called the director. 
Its singularities are topological defects called disclinations. 
Many studies on nematic disclinations include, for instance, the generation of defects followed by the ordering process \cite{Chuang.etal-S1991}, defects' interaction with microparticles \cite{Musevic-Materials2018, Smalyukh.etal-Mol.Cryst.Liq.Cryst.2006, Jiang.etal-Proc.Natl.Acad.Sci.2023,Wang.etal-NC2024} or light \cite{Smalyukh.etal-Mol.Cryst.Liq.Cryst.2006, Meng.etal-NM2023}, operation of molecules by defects \cite{Wang.etal-NatureMater2016, Ohzono.etal-SR2016}, and control of defects by alignment of the surface of the containers \cite{Jiang.etal-Proc.Natl.Acad.Sci.2023, Murray.etal-Phys.Rev.E2014, Yoshida.etal-NatCommun2015, Jiang.etal-Proc.Natl.Acad.Sci.2022,Wang.etal-NC2024,Long.etal-PRX2024}. 
In recent years, nematic ordering and patterns have also been found in living systems \cite{Doostmohammadi.etal-NatCommun2018, Doostmohammadi.Ladoux-TrendsinCellBiology2022}, and topological defects are suggested to be related to some biological functions, including cell extrusion \cite{Saw.etal-N2017}, promotion of bacterial colonies' vertical growth \cite{Copenhagen.etal-NP2021,Shimaya.Takeuchi-PNASNexus2022}, and organization center of \textit{Hydra} morphogenesis \cite{Maroudas-Sacks.etal-NP2020}.
As for theoretical approaches to disclination dynamics, although the governing equations of liquid crystal are well known \cite{deGennes.Prost-Book1995}, it is not straightforward to obtain equations of motion of defects, especially for fully three-dimensional (3D) configurations.
Recently, there has been some theoretical work to better understand the interaction and velocity of 3D topological defects \cite{Long.etal-SM2021, Schimming.Vinals-SM2022, Schimming.Vinals-PRSA2023,Long.etal-PRX2024}.

Despite the intensive study of nematic disclinations, previous observations of disclination dynamics are mainly by transmitted light, and only two-dimensional (2D) information can be obtained.
In our previous work \cite{Zushi.Takeuchi-PNAS2022}, making use of the dye localization at the disclination core \cite{Ohzono.etal-SR2016} to visualize disclination lines, we successfully observed its 3D dynamics, in particular reconnections and loop shrinkage \cite{Zushi.Takeuchi-PNAS2022}. However, the analysis was limited to the dynamics of reconnections occurring to essentially parallel disclinations in an approximately single plane.

In this paper, we study the dynamics of intersecting reconnections, in which two disclinations lying in separate planes approach not in parallel but with an angle. We used confocal microscopy and observed disclinations relaxing from an electrically driven turbulent state. We extracted the positions of disclinations from acquired images, investigated the time evolution of the distance and the angle between two reconnecting disclinations, and discussed the result by comparing it with theory and numerics.

\section{Experimental method}

\begin{figure}
\centering
\includegraphics[width=\hsize]{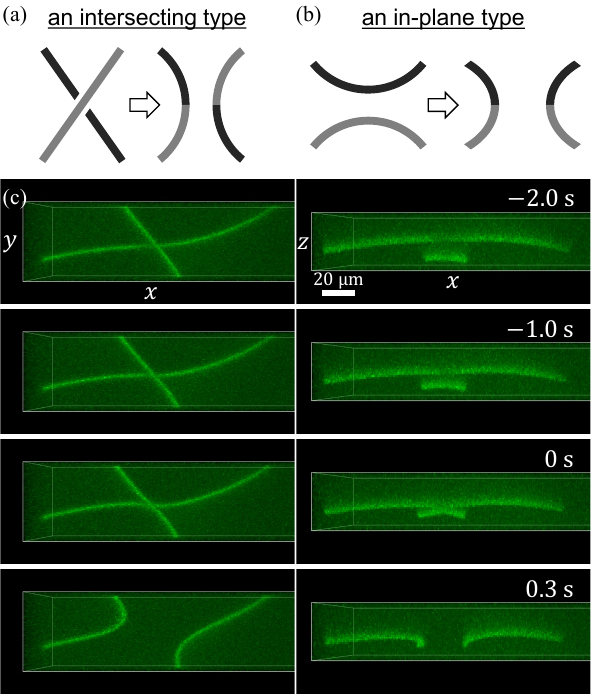}
\caption{
Reconnections of nematic disclinations.
(a,b) Schematics of an intersecting reconnection (a) and an in-plane reconnection (b).
(c) Fluorescence observation of an intersecting reconnection. Top views (left column) and side views (right column) are displayed. The reconnection occurred at time $t=0$. See also Supplemental Videos 1-2 \cite{suppl}. Another reconnection event is also shown in Supplemental Videos 3-4 \cite{suppl}.
}
\label{fig1}
\end{figure}

We used a nematogen MLC-2037 (Merck, a discontinued product) because of small optical anisotropy ($\Delta n = 0.0649$) for optical observation and negative dielectric anisotropy ($\Delta \epsilon = -3.1$) for generating disclinations, as described below. We added $0.01 \unit{wt\%}$ of electrolyte, tetra-\textit{n}-butylammonium bromide, to increase its conductivity and $0.005 \unit{wt\%}$ of a fluorescent dye, coumarin 545T, to label disclinations \cite{Ohzono.etal-SR2016, Zushi.Takeuchi-PNAS2022}. The sample was put into a $130\unit{\mu{}m}$-thick cell consisting of a coverslip and a glass plate with polyimide tape spacers. 
Both glass substrates were coated with indium tin oxide and imposed unidirectional planar alignment by polyvinyl alcohol coating and rubbing with velvet cloth. 
Similarly to our previous work \cite{Zushi.Takeuchi-PNAS2022}, by applying an AC voltage of root-mean-square amplitude $150\unit{V}$ and frequency $35\unit{Hz}$, we induced a turbulent state called the dynamic scattering mode 2 \cite{deGennes.Prost-Book1995, Kai.Zimmermann-PTPS1989} to generate a high density of disclinations. Then, by switching off the applied voltage, the system exhibited a relaxation process,
in which disclinations interacted, shrinked, and disappeared. 

To obtain 3D dynamics of disclinations, we used a laser-scanning confocal microscope, Leica SP8 (objective 40x, NA 1.30, oil immersion) with resonant scanner working at $8\unit{kHz}$ and a piezo objective scanner. 
After starting relaxation, we first searched disclination pairs likely to reconnect by moving the field of view, which was fixed thereafter. 
Due to the imposed homogeneous alignment at the cell surfaces, disclinations moved away from the surfaces and were often found near the midplane of the cell thickness. In such a circumstance, reconnection events are conveniently classified to two types: an in-plane reconnection, in which two reconnecting disclinations are almost in a single plane and approach approximately in parallel [\figpref{fig1}{b}], and an intersecting reconnection, in which disclinations are not in a single plane and form a finite angle [\figpref{fig1}{a}]. Here, we observed twelve intersecting reconnection events.

A series of experimental snapshots is shown in \figpref{fig1}{c} (see also Supplemental Videos 1-4 \cite{suppl}).
The fluorescent dye was excited at $488\unit{nm}$ by laser light polarized in the direction perpendicular to the nematic easy axis.
The directions of the laser polarization and the nematic easy axis are denoted by the $x$- and $y$-axes, respectively.
The fluorescence signal in the range between $500$ and $600\unit{nm}$ was confocally detected by a photomultiplier tube detector (pinhole size $23\unit{\mu m}$). The voxel size in the $xy$-plane was $0.455\unit{\mu m}$ and the spacing between $z$ slices was $1\unit{\mu m}$. The number of voxels was $512, 96, 30$ in the $x$-,$y$-, and $z$-directions, respectively. The time interval between consecutive confocal images was $0.330\unit{s}$. We extracted the coordinates from the 3D $xyz$ images by applying the snake method \cite{Kass.etal-IntJComputVision1988} (see Appendix\,\ref{sec:ImageAnalysis}).  

\section{Experimental results}

\begin{figure}
    \centering
    \includegraphics[width=\hsize]{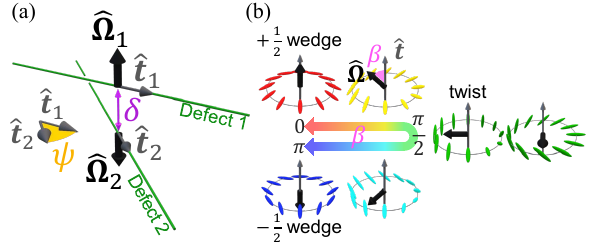}
    \caption{Schematics of two reconnecting disclinations and their director structures. 
    (a) Two disclinations with the respective tangent vectors $\vu{t}_1, \vu{t}_2$ and the rotation axes $\vu{\Omega}_1, \vu{\Omega}_2$ are shown. The two disclinations separate a minimum distance $\delta$ and form an angle $\psi = \arccos( \vt_1 \cdot \vt_2) \leq \pi/2$ with the tangent vectors at the two closest points. 
    (b) Various director configurations of 3D disclination lines, seen at a cross-section perpendicular to the line. All these are topologically equivalent, i.e., homeomorphic. The thick arrows indicate the rotation axis $\vu{\Omega}$, a unit vector normal to the plane in which the director rotates by $\pi$ along a closed loop around the disclination.}
    \label{fig:schematic}
\end{figure}

\begin{figure}
\centering
\includegraphics[width=\hsize]{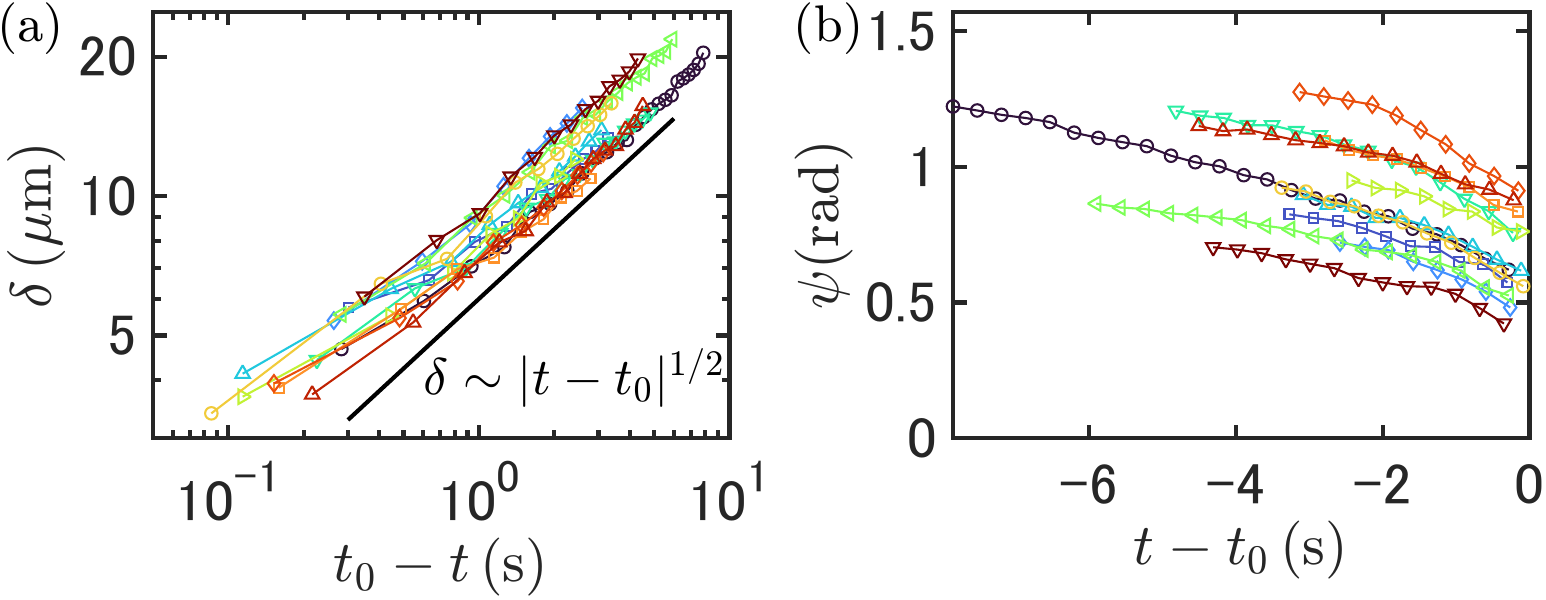}
\caption{
Time dependence of the distance $\delta$ (a) and the angle $\psi$ (b) between two reconnecting disclinations. Results for all 12 reconnection events are shown with different colors.
}
\label{fig2}
\end{figure}

We start with the minimum distance $\delta \qty(t)$ between two reconnecting disclinations at time $t$ [\figpref{fig:schematic}{a}]. The result in \figpref{fig2}{a} shows the square-root law:
\begin{equation}
    \delta \qty(t) \simeq C \qty|t-t_0|^{1/2},  \label{eq:SquareRootLaw}
\end{equation}
where $t_0$ is the reconnection time. This scaling is the same as that observed for in-plane reconnections of essentially parallel disclination pairs \cite{Zushi.Takeuchi-PNAS2022}.
Theoretically, the scaling was derived from the effective force inversely proportional to the distance $\delta$ \cite{Long.etal-SM2021} under the assumption that the force is balanced to a drag force proportional to the disclination velocity. It was also predicted by the disclination density tensor method \cite{Schimming.Vinals-PRSA2023} calculated from the nematic tensor $\vb{Q}$, which contains the director $\vu{n}$ and the scalar order parameter, 
and observed in simulations \cite{Schimming.Vinals-SM2022}. 
From the viewpoint of universality, we note that a similar scaling was also observed for quantum vortices in superfluid helium 4
\cite{Bewley.etal-PNAS2008, Paoletti.etal-PRL2008, Paoletti.etal-PhysicaD2010, Fonda.etal-PNAS2014, Minowa.etal-SA2022} and in the Gross-Pitaevskii equation \cite{Villois.etal-Phys.Rev.Fluids2017}.

For intersecting reconnections, the angle between the two disclinations is another quantity characterizing them. 
We consider the two closest points of the reconnecting disclinations 1 and 2, with tangent vectors $\vt_1$ and $\vt_2$ for the respective disclinations [\figpref{fig:schematic}{a}]. 
We define the inter-disclination angle $\psi$ by $\psi = \arccos( \vt_1 \cdot \vt_2 )$, as shown in the inset of \figpref{fig:schematic}{a}. 
Since the sign of $\vt_1$ and $\vt_2$ is arbitrary, we choose it in such a way that $\vt_1 \cdot \vt_2 \geq 0$, i.e., $0 \leq \psi \leq \pi/2$. 
Our analysis shows that the angle $\psi$ is not constant during each reconnection event but decreases over time [\figpref{fig2}{b}]; in other words, each disclination pair tends to be closer to parallel as time goes on. 
This behavior is consistent with earlier numerical observations \cite{Schimming.Vinals-SM2022, Schimming.Vinals-PRSA2023} and theoretical prediction \cite{Schimming.Vinals-PRSA2023}.

\begin{figure}
\centering
\includegraphics[width=\hsize]{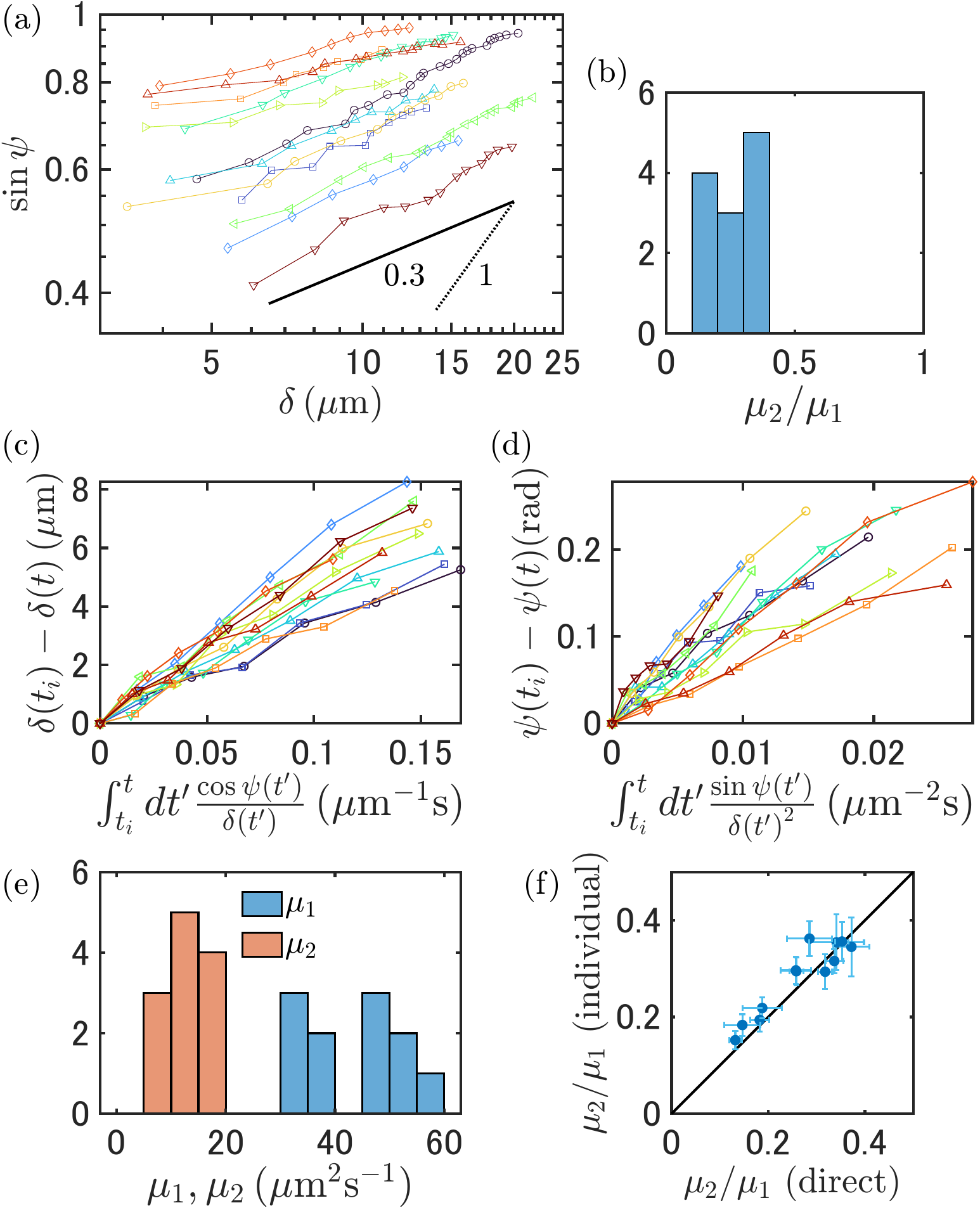}
\caption{
Comparison with the theory.
(a) Log-log plot of $\sin \psi$ and $\delta$.
(b) Histogram of the coefficient ratio $\mu_2 / \mu_1$. The average value over events is $\mu_2 /\mu_1=0.26\pm0.08$ (the error indicates the standard deviation).
(c,d) Evaluation of each of the two equations, \eqref{eq:mu1-int} (c) and \eqref{eq:mu2-int} (d). Images from eight to two time frames before the reconnection were used for all events.
For (a), (c), and (d), results for all 12 reconnection events are shown with different colors.
(e) Histograms of the respective coefficients $\mu_1$ and $\mu_2$. The values are obtained by the fitting of the data in (c) and (d).
(f) Scatter plot of the coefficient ratio obtained from the individual values of $\mu_1$ and $\mu_2$ in (e) against that directly obtained from the power law in (a). Error bars indicate the $95\%$ confidence intervals evaluated from the respective fittings.
}
\label{fig3}
\end{figure}

Let us quantitatively compare the time evolution of the distance $\delta$ and the angle $\psi$, which we obtained experimentally, with the theory reported in Ref.~\cite{Schimming.Vinals-PRSA2023}. 
According to the theory, the time evolution of the distance $\delta$ and the angle $\psi$ are described by the following non-dimensionalized simultaneous differential equations:
\begin{align}
    \dv{\delta}{t} &= 4\qty( \vO_1 \cdot \vO_2 )\frac{\cos\psi}{\delta}, \label{eq:Rpsi-R}\\
    \dv{\psi}{t} &= 4\qty( \vO_1 \cdot \vO_2 )\frac{\sin\psi}{\delta^2}. \label{eq:Rpsi-psi}
\end{align}
Here, the unit vectors $\vu{\Omega}_1$ and $\vu{\Omega}_2$ are the rotation axes, each normal to the plane in which the director rotates by $\pi$ along a closed loop around the disclination [\figpref{fig:schematic}{b}].
Dividing \eqref{eq:Rpsi-R} by \eqref{eq:Rpsi-psi} 
and integrating the result, we obtain
\begin{equation}
    \sin \psi \propto \delta. \label{eq:prop-Rpsi}
\end{equation}
This relation is tested with the experimental data in \figpref{fig3}{a}. 
The result shows that, while $\delta$ and $\sin \psi$ are indeed related by a power law, the exponent value was estimated at $0.26\pm0.08$ by averaging over all events (the error indicates the standard deviation), clearly smaller than 1, the prediction of \eqref{eq:prop-Rpsi}.
This indicates a quantitative limitation of Eqs.\,(\ref{eq:Rpsi-R}) and (\ref{eq:Rpsi-psi}).

To probe the dynamics of the disclinations, suppose Eqs.\,(\ref{eq:Rpsi-R}) and (\ref{eq:Rpsi-psi}) are replaced by the following more general form, with two different coefficients $-\mu_1, -\mu_2 <0$ instead of the common one, $4(\vO_1 \cdot \vO_2)$ in Eqs.\,(\ref{eq:Rpsi-R}) and (\ref{eq:Rpsi-psi}):
\begin{align}
    \dv{\delta}{t} &= -\mu_1 \frac{\cos\psi}{\delta}, \label{eq:mu1}\\
    \dv{\psi}{t} &= -\mu_2 \frac{\sin\psi}{\delta^2}. \label{eq:mu2}
\end{align}
The coefficients may be thought of as effective mobilities for the time evolution of $\delta$ and $\psi$.
We remind that we chose the signs of $\vt_1$ and $\vt_2$ that satisfy $\cos \psi \geq 0$; as a result, $\vO_1 \cdot \vO_2 <0$ is expected for an attracting disclination pair \cite{Long.etal-SM2021, Schimming.Vinals-SM2022}. This is why we set the negative coefficients $-\mu_1, -\mu_2 <0$.
Then, similarly to the derivation of \eqref{eq:prop-Rpsi}, by dividing \eqref{eq:mu1} by \eqref{eq:mu2} and integrating the result, we obtain
\begin{equation}
    \sin \psi \propto \delta ^{\mu_2/\mu_1}. \label{eq:powerlaw-Rpsi}
\end{equation}
This power law is to compare with the experimental data in \figpref{fig3}{a}.
The values of the power-law exponent, $\mu_2/\mu_1$, varied a little among reconnection events [\figpref{fig3}{b}], with the average being $\mu_2 /\mu_1 = 0.26\pm0.08$ as already noted.

Equations (\ref{eq:mu1}) and (\ref{eq:mu2}) were also tested respectively. Since derivatives of experimental data are noisy, not Eqs.\,(\ref{eq:mu1}) and (\ref{eq:mu2}) themselves but the following integrated forms are considered: 
\begin{align}
    \delta \qty(t_\mathrm{i}) - \delta \qty(t) &= \mu_1 \int ^{t}_{t_\mathrm{i}} \dd{t^{\prime}} \frac{\cos \psi\qty(t^{\prime})}{\delta\qty(t^{\prime})}, \label{eq:mu1-int} \\
    \psi   \qty(t_\mathrm{i}) - \psi   \qty(t) &= \mu_2 \int ^{t}_{t_\mathrm{i}} \dd{t^{\prime}} \frac{\sin \psi\qty(t^{\prime})}{\delta\qty(t^{\prime})^2}. \label{eq:mu2-int}
\end{align}
Here, $t_\mathrm{i}$ is set to be the first frame used in the analysis.
By using images from eight to two time frames before the reconnection, we indeed confirm that the l.h.s is proportional to the integral in the r.h.s. for both of \eqsref{eq:mu1-int} and \pref{eq:mu2-int} [\figpref{fig3}{c,d}].
This indicates that Eqs.\,(\ref{eq:mu1}) and (\ref{eq:mu2}) describe intersecting reconnections.
Moreover, from the proportionality coefficients, the individual values of $\mu_1$ and $\mu_2$ are obtained [\figpref{fig3}{e}]. 
For each event, the coefficient ratio is consistent with the value obtained directly from \eqref{eq:powerlaw-Rpsi} [\figpref{fig3}{f}].

To summarize the results so far, Eqs.\,(\ref{eq:mu1}) and (\ref{eq:mu2}) successfully describe the evolution of the distance and the angle of intersecting reconnections, and the mobility ratio $\mu_2 / \mu_1 = 0.26\pm0.08$ was obtained.
Physically, this mobility ratio represents the degree to which the angle between disclinations changes with respect to the distance.
The inequality $\mu_2 / \mu_1 < 1$ indicates that the angle between disclinations changes relatively slower than the distance.
However, the theory of Ref.~\cite{Schimming.Vinals-PRSA2023} predicts Eqs.\,(\ref{eq:Rpsi-R}) and (\ref{eq:Rpsi-psi}) with the common coefficient, hence $\mu_2 /\mu_1 = 1$, disagreeing with the experimental result. 
What then determines the mobility ratio $\mu_2 / \mu_1$ in the experiment?

\section{Theory}

To investigate the discrepancy in the mobility ratio $\mu_2 / \mu_1$ between the experimental data and the theory of Ref.~\cite{Schimming.Vinals-PRSA2023},    
we consider here the case of unequal nematic elastic constants. Nematic distortions may be described by the elastic energy density 
\begin{equation} \label{eq:FrankOseen}
    \frac{K_1}{2}\left(\nabla \cdot \mathbf{\hat{n}}\right)^2 + \frac{K_2}{2}\left(\mathbf{\hat{n}}\cdot \left(\nabla \times \mathbf{\hat{n}}\right)\right)^2 + \frac{K_3}{2}\left| \mathbf{\hat{n}} \times \left(\nabla \times \mathbf{\hat{n}}\right)\right|^2
\end{equation}
where $K_1$, $K_2$, and $K_3$ are the elastic constants for splay, twist, and bend, respectively. In the theory derived in Ref.~\cite{Schimming.Vinals-PRSA2023}, it was assumed $K_1$, $K_2$, and $K_3$ were equal; however, in the experimental system this is not the case. MLC-2037, the mesogen used in this experiment, has a smaller twist elastic constant, $K_2 = 6.1 \pm 0.5\unit{pN}$, than splay, $K_1 = 11.6\unit{pN}$, and bend, $K_3 = 13.2\unit{pN}$ \cite{Zushi.Takeuchi-PNAS2022}. 
Therefore, to more closely align the theory with experiments, we derive equations akin to Eqs.\,(\ref{eq:Rpsi-R}) and (\ref{eq:Rpsi-psi})
but for a generalized case of $K_1 = K_3 \neq K_2$. 

We first remark the director structure around a straight line disclination in its normal plane for the case $K_1 = K_3 = K \neq K_2$, discussed already in the literature, e.g., in Ref.\,\cite{Geurst.etal-JPP1975}.
For concreteness, we assume the tangent vector $\vt = \vu{y}$, the rotation vector $\vu{\Omega} = \vu{z}$, and the director around the disclination is given by $\vu{n} = \cos\theta \vu{x} + \sin\theta \vu{y}$ where $\theta = \theta(x,z)$ is a yet undetermined function.
Substituting this into \eqref{eq:FrankOseen} and taking a variational derivative yields the following equation for $\theta$:
\begin{equation} \label{eq:DirectorELEquation}
    K \frac{\partial^2 \theta}{\partial x^2} + K_2 \frac{\partial^2 \theta}{\partial z^2} = 0.
\end{equation}
A solution that accommodates a defect is given by
\begin{equation} \label{eq:DirectorTheta}
    \theta(x,z) = m \arctan\left(\frac{x}{\sqrt{\varepsilon}z}\right),
\end{equation}
where $m$ is a half-integer multiple winding number and $\varepsilon = K / K_2$ is the ratio of elastic constants.
We note that since \eqref{eq:DirectorELEquation} is linear, the full director solution with multiple defects is a superposition of \eqref{eq:DirectorTheta} for each defect.

To derive the equations for $\delta$ and $\psi$, we use the methods of Ref.~\cite{Schimming.Vinals-PRSA2023} to approximate the structure of the nematic tensor order parameter $\vb{Q}$ at the location of the disclination line.
This approximation is then used to approximate the velocity of the disclination line. To do this, we assume that the dynamics of the $\vb{Q}$ tensor is solely due to free energy relaxation:
\begin{equation} \label{eq:QTensorEvolution}
    \frac{\partial \vb{Q}}{\partial t} = -\frac{1}{\gamma} \left[\frac{\delta F}{\delta \vb{Q}}\right]^{TS}
\end{equation}
where $\gamma$ is a rotational viscosity, $F$ is the free energy of the system, and $[\cdot]^{TS}$ denotes the traceless, symmetric part of a matrix.
The relevant elastic free energy density in terms of $\vb{Q}$ is
\begin{equation} \label{eqn:QElasticFreeEnergy}
    f_e = L_1 |\nabla \vb{Q}|^2 + L_2 |\nabla \cdot \vb{Q}|^2
\end{equation}
where $L_1$ and $L_2$ are elastic constants that may be mapped to $K_i$:
\begin{equation} \label{eqn:LMappings}
    \frac{L_2}{L_1} = 2\left(\frac{K_1}{K_2} - 1\right).
\end{equation}
We note that including the $L_2$ term in \eqref{eqn:QElasticFreeEnergy} does not break the degeneracy $K_1 = K_3$.
Since we only consider relaxational dynamics, the equations we derive do not take into account hydrodynamic effects such as backflow.
We assume the system is composed of two, infinitely long disclinations that remain straight with anti-parallel rotation vectors $\vu{\Omega}_1 \cdot \vu{\Omega}_2 = -1$ aligned along the $z$-axis [\figpref{fig:schematic}{a}].
The resulting equations of motion for $\delta$ and $\psi$ are (see Appendix\,\ref{app:Derivation} for details of the calculation):
\begin{align}
    \gamma \frac{d\delta}{dt} &= - \frac{(4 L_1 + 2 L_2) \cos \psi}{\varepsilon\delta}, \label{eqn:DeltaEvo}\\
    \gamma \frac{d\psi}{dt} &= - \frac{4 L_1 \sin \psi}{\delta^2}. \label{eqn:PsiEvo}
\end{align}
The predicted ratio of effective mobilities $\mu_2 / \mu_1$ is
\begin{equation}
    \frac{\mu_2}{\mu_1} = \frac{2 L_1 \varepsilon}{2L_1 + L_2} = 1,  \label{eq:ratio_mu-K}
\end{equation}
where \eqref{eqn:LMappings} is used in the second equality. 
We thus predict the same mobility coefficient ratio as \eqsref{eq:Rpsi-R} and \pref{eq:Rpsi-psi}, independent of the ratio of the elastic constants. 
We note that this prediction does not take into account deformations of the disclination line, and the resulting director field. 
For a deformed line disclination, \eqref{eq:DirectorTheta} no longer holds, and a more complicated director configuration will be assumed by the system.

\begin{figure}
\includegraphics[width=\hsize,clip]{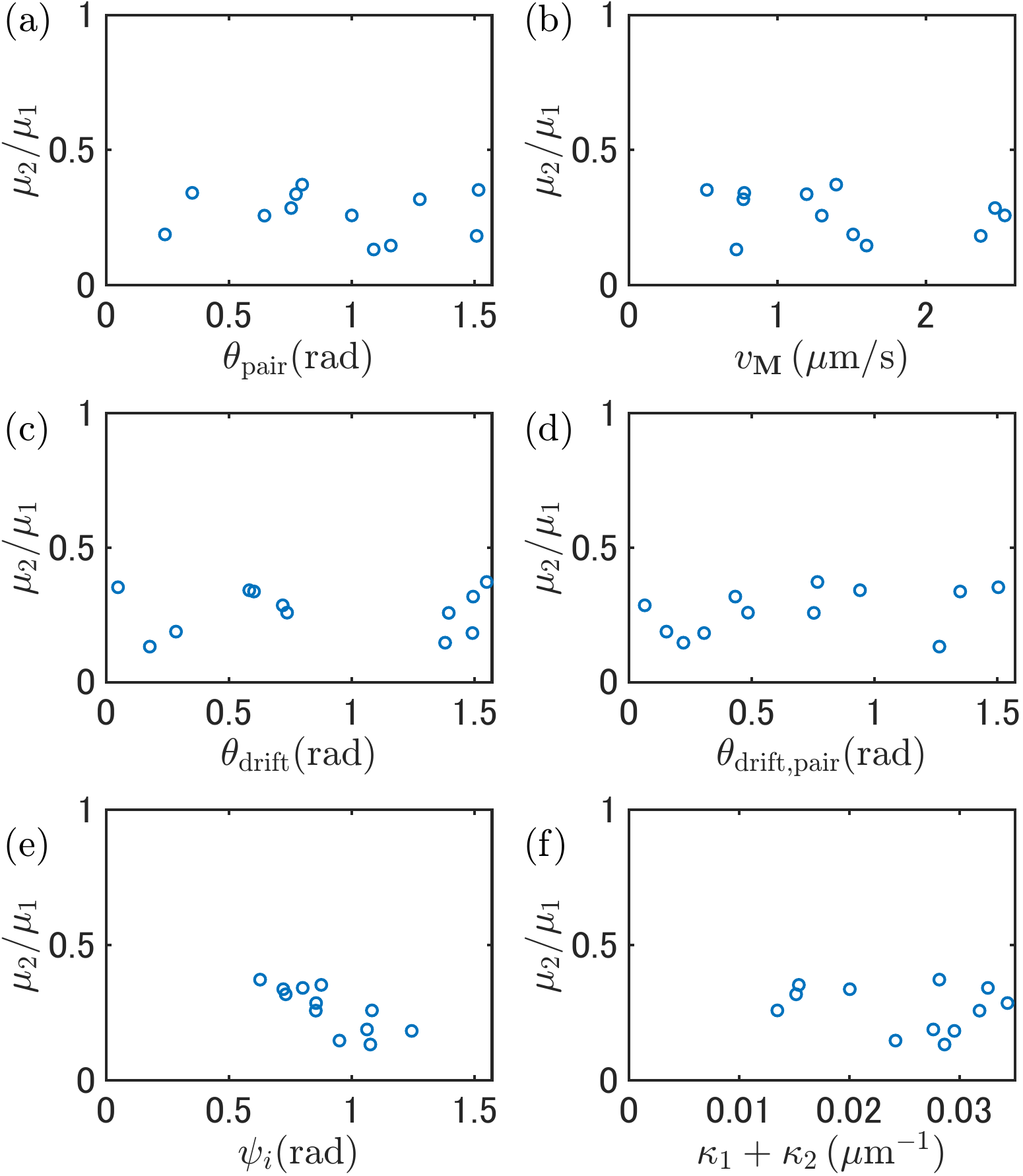}
\centering
\caption{Scatter plots of the coefficient ratio $\mu_2 /\mu_1$ [as determined directly by \eqref{eq:powerlaw-Rpsi}] against several properties of the disclination pairs. 
(a) Against the time-averaged angle of the disclination pair $\theta_\mathrm{pair}$. The angle $\theta_\mathrm{pair}$ is obtained by projecting the vector $\vu{t}_1 + \vu{t}_2$ to the $xy$-plane and measuring the angle between the projected vector and the easy axis ($y$), the direction of the surface alignment of the cell.
(b,c) Against the speed $v_\mathrm{M}$ (b) and the angle $\theta_\mathrm{drift}$ (c) of the drift. 
The drift is a constant flow that the two disclinations are exposed to, which is supposedly induced extrinsically (e.g., by other disclinations present outside the field of view).
We evaluate the drift velocity by using the midpoint of the two closest points. 
Then the angle $\theta_\mathrm{drift}$ is obtained by projecting the drift velocity in the $xy$-plane and measuring the angle it makes with the easy axis.
The drift is also used to define the comoving frame to discuss the symmetry of dynamics in Appendix\,\ref{sec:S-sym}.  
(d) Against the angle $\theta_\mathrm{drift, pair}$ formed by the drift velocity and the disclination pair.
(e) Against the initial angle $\psi_\mathrm{i}$ formed by the two disclinations, which is determined here eight time frames before the reconnection.
(f) Against the curvature of the disclinations. 
The values $\kappa_1$ and $\kappa_2$ are for the two respective disclinations. The curvatures at the closest points right before the reconnection are used.
}
\label{figS-Factors}
\end{figure}

\section{Influencing factors of the mobility ratio $\mu_2/\mu_1$}
\subsection{Experiment}

Since the reduced twist elastic constant alone turned out to be unable to explain the small mobility ratio observed in the experiment, we test if the experimental values of $\mu_2 / \mu_1$ for individual events [\figpref{fig3}{b}] may be correlated with any property of the disclination pairs (\figref{figS-Factors}).
Among the inspected properties, the only significant dependence we found was a slightly negative correlation with the initial angle $\psi_\mathrm{i}$ formed by the two disclination lines [\figpref{figS-Factors}{e}].
This indicates that disclination pairs with larger angles have smaller values of $\mu_2/\mu_1$. Since $\mu_1$ and $\mu_2$ correspond to mobilities for the distance $\delta$ and the angle $\psi$, respectively, we can say that disclinations with larger angles are harder to rotate to be parallel. 
This may suggest the effect of the deformation of disclination lines, induced locally or globally by the rotation of disclinations, which is not considered in the theoretical approach.

\subsection{Simulations}

We also probe the mobility ratio $\mu_2/\mu_1$ by performing 3D computations of the evolution of the $\vb{Q}$-tensor governed by \eqref{eq:QTensorEvolution}, for systems with a reduced twist elastic constant as considered in the previous section.
For the free energy, we use the Maier-Saupe Ball-Majumdar potential for the bulk free energy density \cite{Ball.Majumdar10}, and \eqref{eqn:QElasticFreeEnergy} for the elastic free energy density.
Instead of applying Dirichlet or Neumann conditions, we directly solve the Euler-Lagrange equations for $\vb{Q}$ at the boundary.
The computations are carried out using the Matlab/C++ finite element package FELICITY \cite{Walker18} while matrix inversions are performed using AGMG \cite{Notay10,Napov11,Napov12,Notay12}.
We non-dimensionalize the system by scaling lengths in terms of the nematic correlation length, and times in terms of the nematic relaxation time.
We fix the computational time step to $\Delta t = 0.2$.
Further details involving the computational algorithm can be found in Ref.\,\cite{Schimming.Vinals.Walker21}.

\begin{figure}
\includegraphics[width=\hsize,clip]{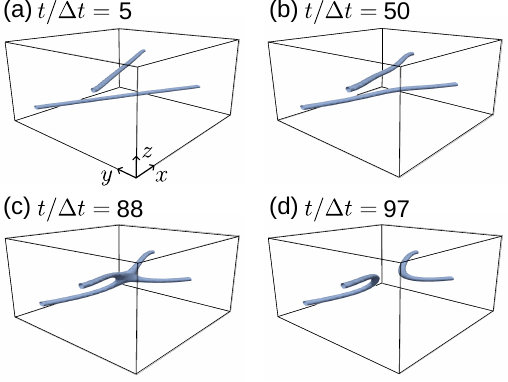}
\centering
\caption{Time snapshots of a simulated disclination reconnection for initial angle $\cos\psi_0 = 0.6$, system size $[L_x,L_y,L_z] = [20,20,10]$, and time steps $t/\Delta t = 5$ (a), $50$ (b), $88$ (c), $97$ (d). The contours indicate points where the nematic scalar order parameter $S = 0.3S_N$, where $S_N$ is the equilibrium value of $S$ in the nematic phase.
}
\label{fig-Snapshots}
\end{figure}

For the computations, we fix the ratio $L_2/L_1 = 2$, corresponding to an elastic constant ratio of $K_2/K_1 = 0.5$, similar to that of the experiment. 
To consider a pair of straight line disclinations that are initially apart by a distance $\delta_0$ in the $z$ direction and form an initial angle $\psi_0$ in the $xy$ plane around the $y$ axis [see also \figpref{fig:schematic}{a}], we initialize the system in a 3D domain of dimensions $[L_x,L_y,L_z]$ with director field
\begin{align}
    \vu{n} &= \cos(\theta_1 - \theta_2)\vu{y} - \sin(\theta_1 - \theta_2)\vu{x}, \\
    \theta_1(x,y,z) &= \frac{1}{2}\arctan\left(\frac{x \cos \psi_0 /2 - y \sin \psi_0/2}{\sqrt{\varepsilon}\left(\delta_0/2 + z\right)}\right), \nonumber \\
    \theta_2(x,y,z) &= \frac{1}{2}\arctan\left(\frac{-x \cos \psi_0/2 - y \sin \psi_0/2}{\sqrt{\varepsilon}\left(\delta_0/2 - z\right)}\right), \nonumber
\end{align}
which satisfies the Euler-Lagrange equation for the director, \eqref{eq:DirectorELEquation}.
We fix the initial distance at $\delta_0 = 4$ and vary $\psi_0$ such that $\cos\psi_0 \in \{0.3,0.4,0.5,0.6,0.7,0.8\}$.
We compute on two different system sizes $[L_x,L_y,L_z] = [10,10,10]$ and $[20,20,10]$ so that, in the latter size, disclinations are initially twice the length of those in the former.
In \figref{fig-Snapshots} we show several time snapshots of a simulated disclination reconnection where the contours represent the locations of the defects.

\begin{figure}
\includegraphics[width=\hsize,clip]{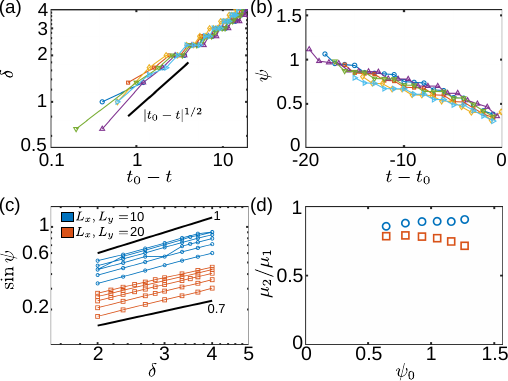}
\centering
\caption{Numerical results for reconnecting disclinations.
(a,b) Time dependence of the distance $\delta$ (a) and the angle $\psi$ (b) between reconnecting disclinations for various simulations with $L_x, L_y = 10$.
(c) Log-log plot of $\sin \psi$ versus $\delta$ for various initial angles $\psi_0$ and two system sizes, $L_x, L_y = 10$ (circles) and $L_x, L_y = 20$ (squares). 
Note that the data for $L_x, L_y = 20$ is vertically offset to distinguish the data sets.
(d) Mobility ratio $\mu_2 / \mu_1$ versus initial disclination angle $\psi_0$ for systems sizes $L_x, L_y = 10$ and $20$.
Lengths and times are given in dimensionless simulation units.
}
\label{fig-Numerics}
\end{figure}

We measure $\delta(t)$ and $\psi(t)$ [\figref{fig-Numerics}(a,b)] and confirm the square-root law \pref{eq:SquareRootLaw} for small $|t-t_0|$, as expected.
The agreement with the square-root law was better for larger system sizes.
Then we plot in \figref{fig-Numerics}(c) the log-log scaling of $\sin\psi$ versus $\delta$ for $\delta \geq 2$ and find a power law relationship between $\sin\psi$ and $\delta$, just as observed in the experiments and predicted by the theory [\eqref{eq:powerlaw-Rpsi}].
For $\delta < 2$ the disclination cores begin to overlap and a clear power law relationship was no longer observed, which is reasonable because the estimation of $\psi$ becomes less accurate and the theory is not expected to hold for strongly overlapping disclination cores.
Although the square-root law \pref{eq:SquareRootLaw} for $\delta(t)$ is not yet established for $\delta \geq 2$, we note that it is not required for the power law \pref{eq:powerlaw-Rpsi} to hold, since theoretically \eqref{eq:powerlaw-Rpsi} holds for all $t$ as long as \eqsref{eq:mu1} and \pref{eq:mu2} are valid, while to obtain \eqref{eq:SquareRootLaw} one needs to take the limit $|t-t_0|\to 0$.

Now, using the power law \pref{eq:powerlaw-Rpsi}, we evaluate the effective mobility ratio $\mu_2 /\mu_1$ by determining the exponent from each computation result shown in \figref{fig-Numerics}(c).
In Fig.~\ref{fig-Numerics}(d) we plot $\mu_2/\mu_1$ as a function of initial angle $\psi_0$ for the set of computations with system size $L_x = L_y = 10$ and $L_x = L_y = 20$. 
For the smaller system size, $\mu_2/\mu_1 \sim 0.9$ and increases slightly as $\psi_0$ increases.
The values are close to the theoretically predicted mobility ratio of $\mu_2 /\mu_1 = 1$, but are smaller, possibly due to the deformations of disclinations as they reconnect.
For the larger system size, in which disclination lengths are doubled, we find that $\mu_2/\mu_1$ is systematically smaller than that of the smaller system size for each $\psi_0$.
Further, $\mu_2/\mu_1$ decreases with increasing $\psi_0$ for the larger system size, similarly to the experiment [\figpref{figS-Factors}{e}].

The numerical results indicate that, for the parameters explored here, the effective mobilities of disclination reconnections are sensitive to large scale deformations of the disclinations. 
The theoretically predicted $\mu_2/\mu_1 = 1$ is most closely assumed by the numerical results at small system sizes, when the disclinations cannot appreciably deform and remain better approximated by straight lines.
When doubling the length of disclinations, the mobility ratio decreases with increasing angle, indicating that disclinations slow their rotation rate due to deformations along the disclination.
These deformations can be seen in \figref{fig-Snapshots} as the disclinations reconnect.

\section{Concluding remarks}

We investigated intersecting reconnections of disclinations. 
Experimentally, the square-root scaling was determined for the time dependence of the distance $\delta$ between reconnecting disclinations. 
The angle $\psi$ between disclinations, which decreases over time, is also important to describe the dynamics. It was found that \eqsref{eq:Rpsi-R} and (\ref{eq:Rpsi-psi}) predicted in Ref.~\cite{Schimming.Vinals-PRSA2023} can describe the experimentally observed time evolution of the distance and the angle, except that the mobility coefficients, $\mu_1$ and $\mu_2$ in \eqsref{eq:mu1} and \pref{eq:mu2}, were found to be different, resulting in a value of the ratio $\mu_2 / \mu_1$ significantly smaller than the theoretical prediction $\mu_2 / \mu_1 = 1$.
We extended the theory to the case of a reduced twist elastic constant and considered the change in the equilibrium director field, but the predicted mobility ratio $\mu_2 / \mu_1$ turned out to remain the same. 
We therefore searched for influencing factors of $\mu_2 / \mu_1$ in both experiments and simulations, and found that $\mu_2 / \mu_1$ tends to decrease with increasing initial angle between the disclination pair. 
Numerically, the mobility ratio was also found to be smaller for longer disclination lines. 
Since disclinations may not rotate while maintaining their straight shapes as assumed in the theory, the rotation results in local deformation of the disclinations, and this effect is stronger when two disclinations are longer or form a larger angle. 
We therefore consider that the deformations of the disclinations may be relevant to the reduced value of $\mu_2 / \mu_1$.

Our results suggest a few interesting directions for future studies. 
First, it is important to develop a theoretical framework to deal with the dynamics of deformable disclinations. 
It may also help to extend the analysis of experimental and numerical data, to analyze not only the vicinity of the closest points but longer parts of the disclinations. 
The surface alignment of the cell may also influence the dynamics of disclinations. 
Second, even though the theory predicted that the mobility ratio $\mu_2/\mu_1$ does not change by a reduced twist elastic constant, previous work \cite{Schimming.Vinals-PRSA2023} has shown that numerical data were closer to theoretical predictions, i.e., \eqsref{eq:Rpsi-R} and \pref{eq:Rpsi-psi} indicating $\mu_2/\mu_1 = 1$, if equal elastic constants were assumed.
This suggests that the reduced twist elastic constant may still have a nontrivial effect, presumably affecting the way the deformation is involved. 
Therefore, it may be interesting to conduct experiments using liquid crystals with different ratios of the elastic constants, $K_2/K_1$, by using, e.g., large $K_2$ expected near the nematic-smectic transition \cite{deGennes.Prost-Book1995} or for nematic discotic liquid crystals \cite{Osipov.Hess-MP1993}.
We also note that deviation of the director field from the equilibrium one [\eqref{eq:DirectorTheta}] can change the mobility ratio $\mu_2/\mu_1$.
For example, if we replace \eqref{eq:DirectorTheta} by that for the one-constant case ($\varepsilon=1$), we obtain $\mu_2/\mu_1 = K_2/K_1 < 1$ in the theory. 
This suggests the potential importance of observing the director field around disclinations, by methods such as the fluorescence confocal polarizing microscopy \cite{Smalyukh.etal-CPL2001,Lavrentovich-PJP2003}, two- or three-photon excitation fluorescence polarizing microscopy \cite{Lee.etal-OL2010, Trivedi.etal-PNAS2012, Ackerman.Smalyukh-PRE2016}, and the tomographic measurement of the dielectric tensor \cite{Shin.etal-NM2022}.

As described in the introduction, since topological defects provide useful means to control microparticles and light in liquid crystal medium, better understanding of defect dynamics can contribute to developments in this direction. 
Moreover, as topological defects also appear in various scientific fields other than liquid crystals, it is also important to unravel the general behavior of topological defects beyond liquid crystals.
We hope our work will contribute to these and trigger further investigations to elucidate fully 3D dynamics of topological defects.

\section*{Author contributions}
K.A.T. designed research. 
Y.Z. performed experiment.
Y.Z. and K.A.T. analyzed data.
C.D.S did theoretical calculation and numerical computations. Y.Z., C.D.S., and K.A.T. wrote the paper.

\begin{acknowledgments}
We thank S. Shankar for his suggestion to use \eqref{eq:powerlaw-Rpsi} to compare the experimental data with the theory.
We acknowledge the material data of MLC-2037 provided by Merck and their permission to present them.
This work is supported in part by Japan Science and Technology
Agency (JST) Precursory Research for Embryonic Science and Technology (PRESTO) (Grant No.~JPMJPR18L6), by KAKENHI from Japan Society for the Promotion of Science (Grant Nos.~JP22J12144, JP22KJ0843, JP19H05800, JP19H05144, JP20H01826, JP23K17664, 24K00593), by JSR Fellowship (The University of Tokyo), and by FoPM, WINGS Program (The University of Tokyo).
C.D.S. acknowledges support from the U.S. Department of Energy through the Los Alamos National Laboratory.
\end{acknowledgments}

\appendix
\section{Image analysis} \label{sec:ImageAnalysis} 

The 3D coordinates of disclinations were extracted from the obtained data of fluorescence intensity at 3D positions $\qty(x,y,z)$ via a method called snakes \cite{Kass.etal-IntJComputVision1988}. It is a way to find smooth contours. Since we know disclinations are smooth lines, the method is suitable for extracting disclinations' positions. In the snake method, the position of a disclination line is represented by $\vb*{v}^\mathrm{} \qty(s) = \qty( x\qty(s), y\qty(s), z\qty(s) )$ with a parameter $s$, and its shape is determined to minimize the total cost function, or energy, given by
\begin{align}
    E^*_\mathrm{} &= \int E_\mathrm{}(\vb*{v}^\mathrm{} \qty(s)) \dd{s}  \notag \\
    &= \int \qty(w_\mathrm{int} E_\mathrm{int}(\vb*{v}^\mathrm{} \qty(s)) + w_\mathrm{im}E_\mathrm{im}(\vb*{v}^\mathrm{} \qty(s)))\dd{s}. \label{eq:5-E_snake}
\end{align}
Here, $E_\mathrm{int }$ is the internal energy, $E_\mathrm{im}$ is the image energy, and $w_\mathrm{int }$ and $w_\mathrm{im}$ are the weights for the internal energy and the image energy, respectively. The internal energy is given by
\begin{equation}
    E_\mathrm{int}  = \frac{1}{2} \qty( \alpha_\mathrm{int} \qty|\pdv{\vb*{v}^\mathrm{}\qty(s)}{s}|^2 + \beta_\mathrm{int} \qty|\pdv[2]{\vb*{v}\mathrm{}\qty(s)}{s}|^2 ), \label{eq:snake-int}
\end{equation}
where $\alpha_\mathrm{int}$ and $\beta_\mathrm{int}$ are coefficients.
The first and the second terms of \eqref{eq:snake-int} correspond to the energy cost due to the length and the roughness of the line, respectively. 
For the image energy, we adopted the intensity $I\qty(x,y,z)$ itself here, since we assume that the fluorescent intensity is higher nearer to the disclination core:
\begin{equation}
    E_\mathrm{im} = -I\qty(x,y,z).
\end{equation}
Practically, the contour $\vb*{v}^\mathrm{}$ was expressed as a series of points, with intervals of approximately $2\unit{\mu m}$, and the minimization of \eqref{eq:5-E_snake} was implemented by the gradient descent method.
Since this implementation can, in principle, only reach local minima, it is important to start from an initial condition close to the desired result. 
For that purpose, disclination shapes roughly estimated from the obtained images were used as initial conditions for the first time frame.
From the second time frame, the result of the previous time frame was used as the initial condition for the subsequent frame. 
In order to reduce noise and improve convergence, the images were filtered with a 3-by-3-by-3 median filter and a Gaussian filter (a kernel with a standard deviation of $0.5$). 
Concerning the time stamp, although the fluorescence intensity at different positions was recorded at slightly different times in the laser scanning confocal microscopy, we used a unique time value for each 3D image, represented by the time at the $z$ coordinate of the midpoint of the two closest points.
The reconnection time $t_0$ was determined not from confocal images but from the 2D image from the transmitted excitation laser, to benefit from the finer time resolution.

\section{Symmetry} \label{sec:S-sym}

\begin{figure}[t!]
\includegraphics[width=0.5\hsize,clip]{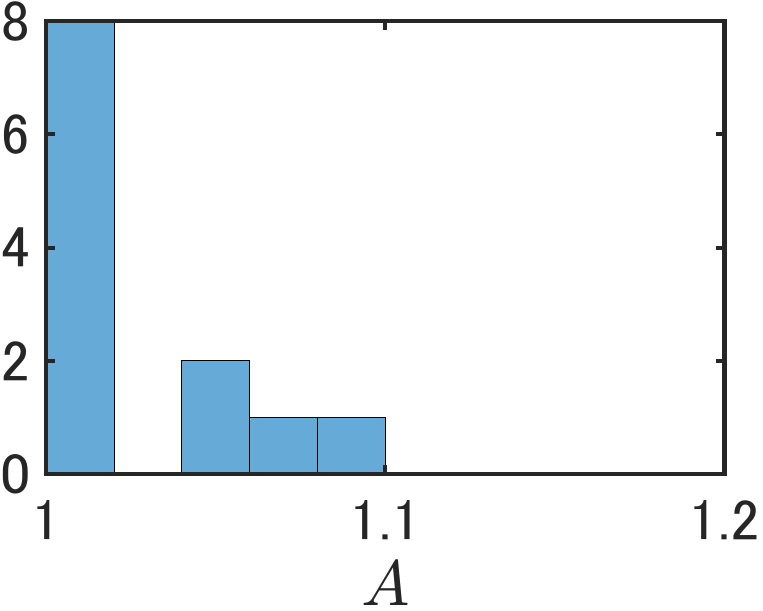}
\centering
\caption{
Histogram of the asymmetry parameter $A$.
}
\label{figS-Asym}
\end{figure}

We discussed the symmetry of defect dynamics in our previous work \cite{Zushi.Takeuchi-PNAS2022}. In 2D, a pair of point-like defects with winding numbers of $\pm 1/2$ is well-known to approach asymmetrically and annihilate, but 3D line-shaped defects were found to approach in a symmetric manner and reconnect, at least for the in-plane reconnections analyzed in Ref.\,\cite{Zushi.Takeuchi-PNAS2022}. This symmetry restoring is explained on the basis of the topological equivalence of various structures of 3D disclinations and the lower energy of twist defects, which result in symmetric dynamics.

Here we inspect the symmetry of the dynamics for the case of intersecting reconnections. As discussed in Ref.\,\cite{Zushi.Takeuchi-PNAS2022}, the disclination dynamics in the lab frame may include extrinsic effects due to other disclinations that exist outside the field of view. Such effects appear as a drift that is uniform in the field of view and has a constant velocity for each reconnection event.
Therefore, these extrinsic effects were removed by evaluating the dynamics in the frame comoving with the constant drift. 
The comoving frame was defined for each reconnection event in a manner similar to that adopted for in-plane reconnections \cite{Zushi.Takeuchi-PNAS2022}. The midpoint of the two closest points of the respective disclination lines was located at each time, and the time evolution of the midpoint was fitted by a linear function of time. The slope of this function determines the constant velocity of the comoving frame.

In the comoving frames, two reconnecting disclinations turn out to approach each other symmetrically for all intersecting reconnections (\figref{figS-Asym}). 
Therefore, the symmetry restoring mechanism discussed in Ref.\,\cite{Zushi.Takeuchi-PNAS2022} also applies to intersecting reconnections, and disclinations are considered to take a twist configuration during the events. 
Asymmetry parameter $A$ was determined for each reconnection event as follows. First, we obtained the distance $D_1\qty(t)$ and $D_2\qty(t)$ between respective disclinations and the reconnection point $\vb{X}_0$, and fit $D_i^2\qty(t)$ by the equation $D_i^2\qty(t) \simeq C_i^2 \qty(t_0 -t)$ with constants $C_i$. Then, $A$ is defined by
\begin{equation}
    A := \frac{\max\{C_1, C_2\}}{\min\{C_1, C_2\}}.\label{eq:asymmetry}
\end{equation}

Here, the estimate of the reconnection point $\vb{X}_0$ was refined by using the time-dependent coordinates of the disclinations before the reconnection.
Specifically, $\vb{X}_0 = \qty(X_0,Y_0, Z_0)$ was determined in such a way that the scaling $D_i(t) \simeq C_i|t-t_0|^{1/2}$ is satisfied more precisely in a time period before the reconnection. 
This was done by evaluating $D_i(t)$ with the reconnection point varied over six neighboring positions in 3D, fitting it to $D_i(t)^2 = a_i|t-t_0| + b_i$, choosing the direction that minimizes $b_1^2 + b_2^2$, and iterating this to reach the (local) minimum. The point was moved at an interval of 1/5 of the voxel size. The final results were at most $2.5\unit{\mu m}$ away from the first rough estimate, which was located from the series of transmitted and confocal images ($X_0$ and $Y_0$ from the transmitted images, $Z_0$ from the confocal images).

\section{Derivation of Theoretical Equations of Motion}\label{app:Derivation}

Here we give more details leading to the equations of motion of \eqsref{eqn:DeltaEvo} and \pref{eqn:PsiEvo}. As mentioned above, we assume the dynamics are purely relaxational and that the system comprises two infinitely long disclinations at distance $\delta$ and angle $\psi$ with respect to one another. We fix a coordinate system so that one disclination is along the $y$-axis, while the other disclination lies in the $xy$-plane, and that the shortest points between disclinations are along the $z$-axis. We also assume the rotation vectors are $\vu{\Omega}_1 = -\vu{\Omega}_2 = \vu{z}$, as sketched in \figpref{fig:schematic}{a}. Finally, we assume the director structure around a single isolated disclination is given by
\begin{equation}
    \mathbf{\hat{n}} = \cos\theta \vu{x} + \sin\theta \vu{y}
\end{equation}
where $\theta$ is given by \eqref{eq:DirectorTheta}.

Following the methods of Ref.\,\cite{Schimming.Vinals-PRSA2023}, we use the kinematic equation for the velocity of a disclination line:
\begin{equation} \label{eqn:DiscV}
    \mathbf{v} = 2 \left. \frac{\vt \times \left(\vu{\Omega}\cdot \mathbf{g}\right)}{|\mathbf{D}|} \right|_{\mathbf{r} = \mathbf{R}}
\end{equation}
where $\vt$ is the tangent vector of the disclination, $g_{\gamma k} = \varepsilon_{\gamma \mu \nu} \partial_t Q_{\mu \alpha} \partial_k Q_{\nu \alpha}$ is related to the topological current, and $D_{\gamma i} = \varepsilon_{\gamma \mu \nu}\varepsilon_{i k \ell} \partial_k Q_{\mu \alpha} \partial_{\ell} Q_{\nu \alpha}$ and we have assumed summation on repeated indices and simplified notation so that $\partial_k \equiv \partial / \partial x_k$. Note that these quantities need only be computed at the location of the disclination core, $\mathbf{r} = \mathbf{R}$. 

The $\vb{Q}$-tensor dynamics are
\begin{multline} \label{eqn:DtQ}
    \gamma \partial_t Q_{\mu \alpha} = 2L_1\partial_k \partial_k Q_{\mu \alpha} + \\
    L_2\left[\partial_{\alpha} \partial_{k} Q_{\mu k} + \partial_{\mu} \partial_k Q_{\alpha k} - \frac{2}{3}\delta_{\mu \alpha} \partial_{\ell} \partial_k Q_{\ell k}\right].
\end{multline}
Upon substituting \eqref{eqn:DtQ} into \eqref{eqn:DiscV}, we find that the last term of \eqref{eqn:DtQ} does not contribute to disclination motion, while the first two $L_2$ terms give the same contribution, so it suffices to simplify and consider only $\gamma \partial_t Q_{\mu \alpha} = 2L_1 \partial_k \partial_k Q_{\mu \alpha} + 2 L_2 \partial_{\alpha}\partial_k Q_{\mu k}$.
Given the new director structure of \eqref{eq:DirectorTheta}, the linear core approximation for $\vb{Q}$ is
\begin{multline}
    \vb{Q} \approx S_N\left[\frac{1}{6}\vb{I} - \frac{1}{2}\vu{z}\otimes\vu{z} \right. \\ 
    + \left.\frac{\sqrt{\varepsilon} z}{2 a}\left(\vu{x}\otimes\vu{x} - \vu{y}\otimes\vu{y}\right) + \frac{x}{2a}\left(\vu{x}\otimes\vu{y} + \vu{y} \otimes \vu{x}\right)\right]
\end{multline}
where $S_N$ is the magnitude of $\vb{Q}$ in the nematic phase and $a$ is the disclination core radius. Substituting this into \eqsref{eqn:DtQ} and \pref{eqn:DiscV} and taking into account the rotation of the director caused by the other disclination line, the velocity of the first disclination line along the $y$-axis is
\begin{equation}
    \vb{v}_1(y) = \frac{(2L_1 + L_2)\delta \cos\psi}{\varepsilon \delta^2 + y^2 \sin^2\psi}\vu{z} - \frac{2L_1 y \varepsilon \sin \psi}{\varepsilon \delta^2 + y^2 \sin^2\psi}\vu{x}
\end{equation}
A similar equation may be derived for the velocity of the other disclination, $\mathbf{v}_2$.
To obtain equations of motion for $\delta$ and $\psi$, we use the identities
\begin{align}
    \frac{d \delta}{d t} &= \mathbf{\hat{z}}\cdot \left(\mathbf{v}_2(0) - \mathbf{v}_1(0)\right), \\
    -\sin \psi \frac{d \psi}{d t} &= \left.\frac{d\mathbf{v}_2}{dy}\right|_{y = 0}\cdot \vt_1 + \left.\frac{d\mathbf{v}_1}{dy}\right|_{y = 0}\cdot\vt_2,
\end{align}
leading to \eqsref{eqn:DeltaEvo} and \pref{eqn:PsiEvo}.

\section{Supplemental Video Caption}

\begin{description}
\item[Video 1-4]
Fluorescence observation of intersecting reconnections. Supplemental Videos 1 and 2 show the event displayed in \figpref{fig1}{c}, from the top and side, respectively. Supplemental Videos 3 and 4 show another reconnection event, again from the top and side, respectively.
\end{description}

\bibliography{ref}

\end{document}